\documentclass[twocolumn,prl,aps,showpacs,superscriptaddress]{revtex4}
\usepackage{epsfig}
\usepackage{fancyhdr}
\begin{document}
\title{Self-assembly of Nanometer-scale Magnetic Dots with Narrow Size 
Distributions on an Insulating Substrate}
\author{Zheng Gai}
\affiliation{Solid State Division, Oak Ridge National Laboratory, 
Oak Ridge, TN 37831}
\affiliation{Department of 
Physics, Peking University, Beijing, 100871, P.R. China}
\author{Biao Wu}
\affiliation{Department of Physics, The University of Texas, Austin, 
Texas 78712}
\affiliation{Solid State Division, Oak Ridge National Laboratory, 
Oak Ridge, TN 37831}
\author{J.P. Pierce}
\affiliation{Department of Physics and Astronomy, University of Tennessee,
Knoxville,Tennessee 37996}
\affiliation{Solid State Division, Oak Ridge National Laboratory, 
Oak Ridge, TN 37831}
\author{G.A. Farnan}
\affiliation{Solid State Division, Oak Ridge National Laboratory, 
Oak Ridge, TN 37831}
\author{Dajun Shu}
\affiliation{International Center for Quantum Structures, Chinese Academy 
of Sciences, Beijing 100080, P.R. China}
\affiliation{Department of Physics, Nanjing University, Nanjing, 210093, P.R. 
China}
\author{Mu Wang}
\affiliation{Department of Physics, Nanjing University, Nanjing, 210093, 
P.R. China}
\affiliation{International Center for Quantum Structures, Chinese Academy 
of Sciences, Beijing 100080, P.R. China}
\author{Zhenyu Zhang}
\affiliation{Solid State Division, Oak Ridge National Laboratory, 
Oak Ridge, TN 37831}
\affiliation{International Center for Quantum Structures, Chinese Academy 
of Sciences, Beijing 100080, P.R. China}
\author{Jian Shen}
\affiliation{Solid State Division, Oak Ridge National Laboratory, 
Oak Ridge, TN 37831}

\date{\today}
\begin{abstract}
The self-assembly of iron dots on the insulating surface of NaCl(001) is investigated 
experimentally and theoretically. Under proper growth conditions,
nanometer-scale magnetic iron dots with remarkably narrow size distributions can
be achieved in the absence of a wetting layer.  Furthermore, both the vertical and 
lateral sizes of the dots can be tuned with the iron dosage without introducing 
apparent size broadening, even though the clustering is clearly in the strong 
coarsening regime. These observations are interpreted using a phenomenological
mean-field theory, in which a coverage-dependent optimal dot size is selected by 
strain-mediated dot-dot interactions.
\end{abstract}

\maketitle
Clustering on surfaces by nucleation and growth during atom deposition has been 
an important subject in basic and applied science for decades
\cite{Venables,Zinke}.  Recent efforts 
have been focused on searching for methods to obtain nanometer-scale clusters with 
narrow size distributions. Such clusters or quantum dots are potentially valuable 
for optical, electronic, and magnetic device applications, but mass production of 
such structures by lithography or etching-based fabrication has proved to be 
exceptionally challenging \cite{Wolf,review}. 
Alternatively, it has been realized that the strain 
energy associated with the lattice mismatch between the dot and the substrate materials 
can be exploited to induce self-assembled formation of quantum dots with narrow size 
distributions. This has generated much excitement, particularly in the area of 
semiconductor quantum dots \cite{Wolf,review,Mo,Jesson,Xie,Xin}. 
In such cases, the growth of the dots often proceeds 
in the Stranski-Krastanow (SK) mode, which is characterized by the presence of a wetting 
layer prior to three-dimensional (3D) clustering. To date, the precise mechanism for size 
selection in semiconductor quantum dot systems remains a subject of active debate 
\cite{review,Mo,Jesson,Daruka}:
some attribute them to strain-induced thermodynamic equilibrium states, while others 
associate them with metastable configurations due to kinetic limitations. 

Although improved size uniformity can be achieved in quantum dot formation via 
the SK growth mode, the presence of a wetting layer is often undesirable, particularly 
for electronic and magnetic device applications of metallic/magnetic quantum dots. For 
this reason, it is preferred to fabricate quantum dots in the Volmer-Weber (VW) growth 
mode, which is characterized by immediate 3D clustering on the substrate surface. Indeed, 
considerable recent efforts have been devoted to metallic/magnetic quantum dot formation 
on various substrates \cite{Fruchart,huang,Baumer}, but no significant size uniformity 
has been achieved in such studies.

In this Letter we investigate the self-assembly of iron dots on NaCl(001), 
an insulating substrate, by thermal deposition and variable-temperature atomic 
force microscopy in ultrahigh vacuum. We show that, by properly choosing 
the growth conditions, nanometer-scale magnetic iron dots with remarkably narrow size 
distributions can be achieved in the absence of a wetting layer (VW growth).  
Moreover, by changing the dosage of iron, we can tune both the vertical and lateral 
sizes of the dots without introducing apparent size broadening, even though the clustering 
is already in the strong coarsening regime, signified by the decrease in dot density 
as a function of the iron dosage. The preserved narrowness in the island size 
distributions is in clear contradiction with the expectations of existing understanding 
of clustering on surfaces in the coarsening regime \cite{Zinke}. 
We interpret these observations 
within a phenomenological mean-field theory, in which a coverage-dependent optimal dot 
size is selected by the competition between the self-energy of a dot, i.e. the total 
energy of an isolated Fe dot on NaCl(100), and the energy of the strain-induced dipolar 
interaction between the dots. 

The experiments were performed in an ultrahigh vacuum (UHV) system with base pressure 
of 1$\times$10$^{10}$ torr. The system is equipped with electron beam sources, laser 
molecular beam epitaxy, and an \textit{in-situ} Omicron variable-temperature UHV beam 
deflection atomic force microscope (AFM)/ scanning tunneling microscope (STM) with 
cooling and heating facilities covering a temperature range of (13-1500) K. The 
noncontact mode AFM was used to study the surface morphology in this work. The NaCl 
single crystal substrates were cleaved in air, then were immediately loaded into the 
UHV chamber and were annealed to 530K for one hour to remove surface contamination 
prior to the experiments. AFM images of the cleaved NaCl(001) surface show monatomic 
height steps and large terraces with no detectable adsorbates. The iron was evaporated 
from an Fe wire (5N purity) heated by electron beam bombardment at a rate of 
0.04 ML/min (1ML is equivalent to the nominal surface atomic density of bcc Fe(110), 
1.7$\times$10$^{15}$ atoms/cm$^2$). The iron dots can be formed only within a 
finite temperature window, since it is known that the morphology of the substrate 
changes dramatically if the temperature is above 720K \cite{Szymonski}, 
whereas at low temperatures, only random clusters or percolated iron films form \cite{Gai}.
The dots discussed in this 
Letter are grown at a substrate temperature of 530K.

\begin{figure}[!htb]
\begin{center}
\resizebox *{7.0cm}{7.0cm}{\includegraphics*{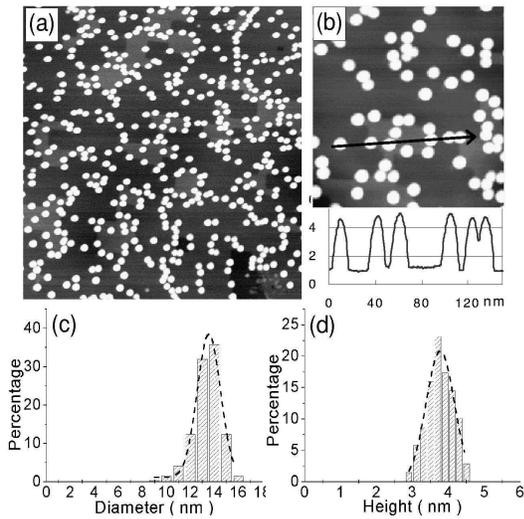}}
\end{center}
\caption{Nanometer-scale iron dots grown on NaCl(001) with a nominal Fe dose of 1.7 ML. 
(a) A typical NC-AFM image. Scan area 500 nm$\times$500 nm. (b) Close up and line profile 
(150 nm long) of the dots. Scan area 200 nm$\times$200 nm. The unit in the line 
profile is nm. (c) and (d) Diameter and height distributions of the dots shown in (a) 
and the corresponding Gaussian fits. The center 
and dispersion for the diameter distribution 
are 13.5 nm and 1.1 nm, respectively; and for the height distribution are 3.76 nm 
and 0.36 nm, respectively.}
\label{fig1}
\end{figure}
Nanometer-scale dots can be formed directly on the NaCl(001) surface without a 
wetting layer, indicating that the growth proceeds in the Volmer-Weber mode. As 
shown in the AFM images in Figs. 1 and 2, deposited Fe atoms tend to nucleate and 
form dots even at submonolayer coverage. The preferred VW growth mode is not too 
surprising, given the large surface free energy difference (2.48 J/m$^{2}$ for 
Fe versus 0.18 J/m$^{2}$for NaCl)\cite{Boer} and the large lattice mismatch between Fe 
and NaCl\cite{ltt}. The most eye-catching characteristic of the dots is their size 
uniformity. Fig. 1 (a) shows a representative AFM image of the dots with a nominal 
Fe thickness of 1.7ML. The very uniform dots are randomly distributed on the terraces. 
Fig. 1 (b) is a close up together with a line profile of the dots. Estimated from the 
line scan, the dots are 3.5 to 4 nm in height, and around 14 nm in diameter \cite{ldia}. 
Fig. 1 (c) and (d) are the lateral size and height distributions of the dots shown 
in Fig.1 (a) along with the corresponding Gaussian fits. Consistent with the line 
profile, the centers of the Gaussian fits of the height and diameter distributions of 
the dots appear at 3.76 nm and 13.5 nm, respectively. Each distribution has a very narrow 
width. The dispersions, $\Delta h=(\langle h^2\rangle-\langle h\rangle^2)^{1/2}$, 
of the height and diameter distributions are only 0.36 nm and 
1.1 nm, respectively, which are less than 10\% of the average height and diameter values. 
Such narrow dispersions are comparable with the narrowest dispersions achieved in 
semiconductor quantum dot growth via the SK mode \cite{Xin}, and are particularly 
striking because clusters grown via the VW mode normally have much broader distributions 
\cite{Wolf,Fruchart,huang,Baumer}.

\begin{figure}[!htb]
\begin{center}
\resizebox *{7.0cm}{11.0cm}{\includegraphics*{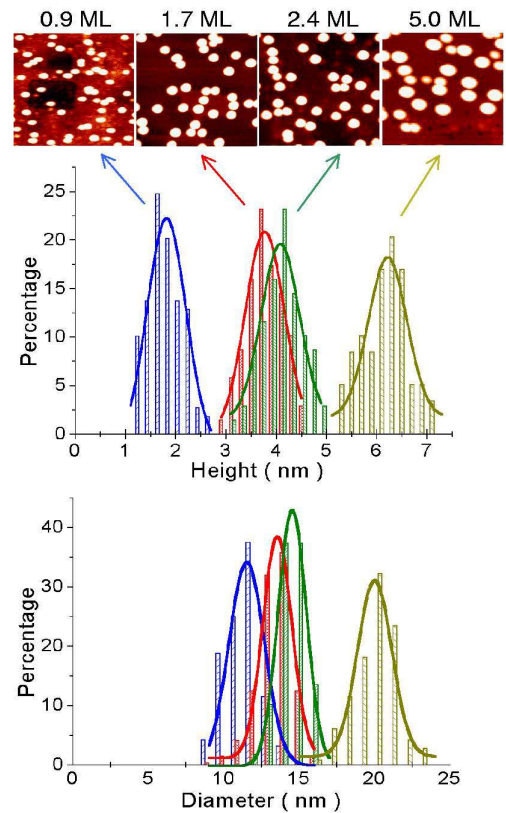}}
\end{center}
\caption{Typical NC-AFM images and height and diameter distributions for dots grown 
on NaCl(001) at different Fe coverages $\theta$. The image area for each 
coverage is 150 nm$\times$150 nm. The dispersions for the height distributions for the four 
coverages are all 0.36nm; and the dispersions for the diameter distributions are 
0.9nm, 1.1nm, 1.1nm, and 1.2nm, for the four coverages, respectively. }
\label{fig2}
\end{figure}
Another striking observation lies in the coverage dependence of the dots shown in 
Fig. 2. As the Fe coverage increases, the dots grow larger and taller, but an optimal
size of the dots is always selected while the narrowness of their lateral and vertical 
size distributions is preserved.  Our quantitative analysis of the dispersions does not 
yield any significant additional (coverage-dependent) broadening of the dot size 
distributions, consistent with the impression drawn by naked eye from the AFM images 
shown in Fig. 2. In Fig. 3, the optimal height and lateral size of the dots are plotted 
as functions of the iron coverage, together with the density of the dots. The decrease 
of the dot density with increasing coverage indicates that the growth of the dots is 
clearly in the coarsening regime. This observation on the 
variation of the island density makes the persistence of the narrow size distributions 
even more dramatic, because in the coarsening regime, the \textit{absolute} size 
distribution of the islands is expected to broaden with increasing coverage \cite{Zinke}.

 Preliminary surface magneto-optical Kerr effect (SMOKE) measurements show that these 
iron dots are superparamagnetic in nature. The details of the magnetic investigation, 
however, are beyond the scope of the present paper, which focuses on the formation of 
the highly uniform dots. In order to understand the underlying formation mechanism of 
these nanoclusters, it is highly desirable to know both their detailed atomic structures 
and their stability. Unfortunately, the iron dots formed on NaCl cannot be imaged by STM 
with higher (atomic) resolution. On the other hand, \textit{in-situ} time (from immediately
 after formation up to 10 days) and temperature (from room temperature up to 550 K) 
dependent AFM studies reveal little change in the morphology of the Fe dots, indicating 
that the dot arrays on the surface are in an energetically favorable configuration that 
is at least metastable.
\begin{figure}[!htb]
\begin{center}
\resizebox *{5.0cm}{6.0cm}{\includegraphics*{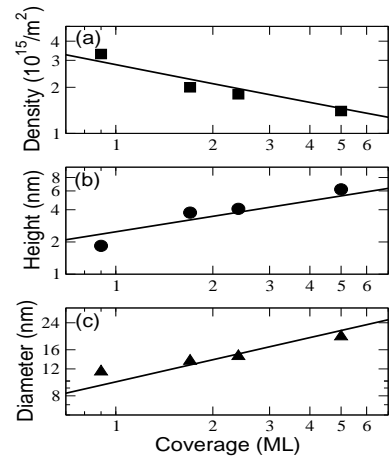}}
\end{center}
\caption{The density (a), the peak height (b), and the peak diameter (c) of 
the iron dots as functions of the Fe coverage obtained from the Gaussian fits, 
in comparison with the mean field theory predictions (straight lines). Only a single 
parameter, $\alpha=0.46$, is used to fit all the slopes of the three independent plots.}
\label{fig3}
\end{figure}

\begin{figure}[!htb]
\begin{center}
\resizebox *{4.0cm}{3.0cm}{\includegraphics*{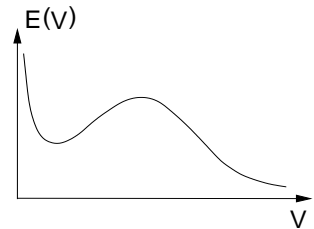}}
\end{center}
\caption{
Schematic drawing of the energy per atom for the Fe dot assembly as a function of 
the volume $V$ at a given coverage, showing the existence of local energy minimum. 
}
\label{fig4}
\end{figure}
At the present, the precise underlying mechanism for the formation of these narrow-sized 
iron dots is not completely clear, but the following phenomenological mean-field theory, 
stressing on strain-mediated dot-dot interactions, is able to reproduce most of the salient 
features observed in the experiment. First, the fact that the preferred size of the Fe 
dots is hardly changed upon annealing clearly indicates that the energy per atom
$E$, equivalently, the chemical potential, of the Fe dot assembly must have a 
local minimum at the observed optimal size for a given coverage. This suggests that the 
qualitative form of the energy $E$ follows the generic behavior shown in Fig. 4 
for a given coverage: When the volume approaches zero, $E$ is close to the 
energy of one adatom; when the dot grows very large, $E$ is about the energy 
per atom in a bulk Fe crystal, which is much smaller than the energy of one adatom. 
Between the two limits there exists a local minimum, mainly induced by the elastic 
relaxation caused by the discontinuity of the intrinsic surface stress tensor at the 
dot edges \cite{review}. Secondly, the observation of different optimal sizes at different 
coverages shown in Fig. 2 suggests that the local energy minimum is a function of 
the iron coverage, implying that dot-dot interactions should contribute substantially 
to the energy $E$. Based on these considerations, we can express $E$ as 
a function of both the iron coverage $\theta$ and the dot size $V$
within the mean field approximation as
\begin{equation}
E(\theta,V)=E_0(V)+P(\theta,V)\,,
\end{equation}
where $E_0$ is the self-energy of an isolated dot and $P$ 
is the dot-dot interaction energy.

The interaction energy per atom,$P(\theta,V)$, is originated from misfit-induced 
strain within 
the system. By adopting the standard dipole-dipole interaction between two dots as
$P(V_1,V_2)\propto V_1V_2/r^3$\cite{review}, we have  
\begin{equation}
\label{eq:p}
P(\theta,V)=g{\theta^{3/2}\over V^{1/2}}\,,
\end{equation}
where $g$ is a constant related to the interaction strength and the effective number of 
the nearest neighbors of a given dot. We note that, on one hand, the long range dot-dot 
interaction is important enough in influencing the energy per atom of the dot assembly, 
thereby the existence of an optimal dot size; on the other hand, it is not strong enough 
to cause the center of mass of a given dot to change, thereby the lacking of the spatial 
order among the dots\cite{Liu}.

The self-energy $E_0(V)$ of an isolated dot of size $V$ includes all possible 
contributions to 
the dot, such as the strain energy, interface energy, step energy, and kink energy, 
but it does not depend on the total coverage \textit{$\theta$}. In principle, 
an explicit, but rather complex, expression can be obtained from existing studies 
of strain-mediated dot formation in heteroepitaxy, such as Eq. (4.15) of 
Ref.\cite{review}. But surprisingly, our detailed analysis based on Eq. (4.15)\cite{review}
shows that no qualitatively correct fitting could be obtained for the range of iron 
coverages explored in the present experiment. In the following, we choose the simple 
expression
\begin{equation}
E_0(V)=bV^{\alpha}\,.
\end{equation}
as an alternative phenomenological approach, and show that the corresponding energy 
per atom  is capable of explaining many of the salient features observed in the experiment. 

First, by setting the first derivative of the energy per atom $E(\theta,V)$ with respect 
to the 
volume $V$ equal to zero, we obtain a coverage-dependent local energy minimum, 
at the volume $b\alpha V^{\alpha+1/2}={g\over 2}\theta^{3/2}$. 
This, in turn, yields the relation between the density of the dots $\theta/V$
and the coverage $\theta$
\begin{equation}
b\alpha\Big({\theta\over V}\Big)^{-(\alpha+1/2)}={g\over 2}\theta^{1-\alpha}\,.
\end{equation}
With $\alpha$ = 0.46, our theory gives an excellent fit to the experimental data 
as shown in Fig. 3(a). Next, assuming that the equilibrium shape of the dots does not 
change with the volume \cite{Duport}, we can also obtain the dot height (\textit{h}) 
and the diameter (\textit{D}) as functions of the coverage, given by 
$h\propto \theta^{0.52}$ and $D\propto \theta^{0.52}$, 
respectively. These two relationships also find very good fits to the experimental 
data as shown in Figs. 3(b) and 3(c). The fact that the exponent $\alpha$ as a single 
fitting parameter yields very good fittings to the slopes of three independent sets 
of experimental observations hints on the physical validity of the choice made in Eq. (3). 
Whereas the precise underlying physics for the success of Eq. (3) remains to be 
explored, we suspect that it could be related to the unusually large lattice mismatch 
in the present system \cite{ltt}.  
Finally, we note that the dispersion of the distribution 
of the island heights (or the diameters) shows little change with the coverage, whereas 
our mean field theory predicts that the dispersion should increase with the coverage,
$\Delta h\propto \theta^{0.4}$. This discrepancy is not too surprising, because 
within the mean-field theory, 
the resulting optimal size of the islands can be reliably obtained, but the exponent of 
the scaling function around the optimal value is typically overestimated, a trend well 
recognized in critical phenomena \cite{Wilson}. 

In summary, we have shown that, under proper growth conditions, the self-assembly of 
iron dots on the insulating surface of NaCl(001) leads to the formation of nanometer-scale
magnetic iron dots with narrow size distributions in the absence of a wetting layer.  
We have also demonstrated that the vertical and lateral sizes of the dots can both
be changed by the iron coverage while the narrow size distributions are preserved, 
even though the clustering is clearly in the strong coarsening regime. 
These striking observations have been interpreted 
successfully with a phenomenological mean-field theory.

We are grateful to E.W. Plummer for invaluable discussions.
This work was supported in part by the LDRD of ORNL, 
managed by UT-Battelle, LLC for the USDOE (
DE-AC05-00OR22725), and by the NSF (DMR-0071893 (BW) 
and DMR-0105232 (JP)).


\begin{thebibliography}{99}
\bibitem{ltt}
The lattice constants of bcc Fe and NaCl are 2.87 $\mathring{\rm A}$ 
and 5.46 $\mathring{\rm A}$,  
the nearest neighbors of bcc Fe(110) and NaCl(100) are separated by 2.49 
$\mathring{\rm A}$ and 3.99 $\mathring{\rm A}$, 
respectively. The Fe dots are likely in bcc phase because its fcc phase exists only in 
bulk at temperatures above 1186 K, or in Fe films grown on fcc substrates 
(e.g. Cu(100)) whose lattice constants are very close to that of Fe.
\bibitem{ldia}
The lateral diameters of the dots are overestimated because of the 
tip effect. However, the lateral size can be estimated with an assumed 
shape of the dots in addition to the known information of height, density, and total 
dosages. For the 1.7 ML Fe dot, assuming a truncated 
pyramid shape, the average lateral size of the dots is around 7 to 8 nm 
instead of 14 nm.
\end{thebibliography}
\end{document}